\documentclass{jnmp}

\newcommand{\RR}{{\mathbb R}}

\usepackage{amsmath}

\JNMPnumberwithin{equation}{section}
\resetfootnoterule

\newtheorem{theorem}{Proposition} 

\newtheorem*{acknowledgments}{Acknowledgments}
\theoremstyle{definition}

\begin{document}
\renewcommand{\evenhead}{A.I. Zenchuk }
\renewcommand{\oddhead}{Combination of  inverse spectral transform method and ... }

\thispagestyle{empty}


\copyrightnote{200*}{A I Zenchuk}

\Name{Combination of  inverse spectral transform method and  method of characteristics: deformed Pohlmeyer equation}

\label{firstpage}

\Author{A.I.Zenchuk}

\Address{ Center of Nonlinear Studies of L.D.Landau Institute
for Theoretical Physics  
(International Institute of Nonlinear Science)
Kosygina 2, Moscow, Russia 119334\\
~~E-mail: zenchuk@itp.ac.ru}

\Date{Received Month *, 200*; Accepted in Revised Form Month *, 200*}

\begin{abstract}
We apply a version of the dressing method  to a system of   four dimensional nonlinear Partial Differential Equations (PDEs), which contains both  Pohlmeyer equation  (i.e. nonlinear PDE integrable by the Inverse Spectral Transform Method)
 and nonlinear matrix PDE integrable by the method of characteristics as particular reductions. 
  Some other reductions are suggested.
\end{abstract}


\section{Introduction}

In this paper we apply a properly modified  version of the dressing method developed  in \cite{Z4} to a system of nonlinear Partial Differential Equations (PDEs) which combines some properties of both nonlinear   PDEs
integrable by the Inverse Spectral Transform Method (ISTM) (or $S$-integrable PDEs) \cite{ZSh1,ZSh2,ZM,BM} and nonlinear PDEs integrable  by the method of characteristics \cite{Whitham,SZ1,ZS2}.
 The important feature of this  version is that it is based on the integral operator with nontrivial kernel
\cite{ZS2,ZS}, unlikely the classical $\bar \partial$-dressing method \cite{ZM,BM,K}. 
The system of nonlinear PDEs studied below can be written as a 
system of evolution equations,
\begin{eqnarray}\label{nl}
&&
w_{t}+u_{x_1} p + w_{x_1} w =0,\\\nonumber
&&
p_{t}+v_{x_1} p + p_{x_1} w =0,\\\nonumber
&&
(u_t - w u_{x_1} - u v_{x_1})_{x_2} =(u_{x_3} - w u_{x_2} - u v_{x_2})_{x_1},\\\nonumber
&&
(v_t - p u_{x_1} - v v_{x_1})_{x_2} =(v_{x_3} - p u_{x_2} - v v_{x_2})_{x_1},
\end{eqnarray}
supplemented by the pair of compatible constraints
\begin{eqnarray}\label{nl_constr}
&&
w_{x_3}+u_{x_2} p + w_{x_2} w =0,\\\nonumber
&&
p_{x_3}+v_{x_2} p + p_{x_2} w =0,
\end{eqnarray}
where fields $w$, $p$, $v$ and $u$ are $Q\times Q$ matrices.
This system reduces to Pohlmeyer equation \cite{P2,P3} (which is $S$-integrable PDE) if $p=u=w=0$,
\begin{eqnarray}\label{P}
v_{tx_2}-v_{x_3x_1} =[v_{x_2}, v_{x_1}]\;\;\;\Leftrightarrow \;\;\;
(J^{-1} J_{t})_{x_2} -(J^{-1} J_{x_3})_{x_1} =0
\end{eqnarray}
(where $v_{x_1}=J^{-1} J_{t}$, $v_{x_2}=J^{-1} J_{x_3}$)
and to to the pair of compatible matrix first order quasilinear PDEs integrable by the method of characteristics if $v=p=u=0$,
\cite{SZ1}:
\begin{eqnarray}
w_{t}+ w_{x_1} w =0,\;\;\;w_{x_3}+ w_{x_2} w =0.
\end{eqnarray}
An interesting reductions of the system (\ref{nl},\ref{nl_constr}) is the following
(1+1)-dimensional system:
\begin{eqnarray}\label{nlred}
&&
w_{t}+u_{x_1} p + w_{x_1} w =0,\;\;\;
p_{t}+v_{x_1} p + p_{x_1} w =0,\\\nonumber
&&
u_{t} - w u_{x_1} - u v_{x_1} =0 ,\;\;\;v_{t} - p u_{x_1} - v v_{x_1} =0.
\end{eqnarray}
A distinguished feature of this system is evident in the case of
scalar fields $w$, $p$, $u$ and $v$ ($Q=1$), when eqs.(\ref{nlred}) read:
\begin{eqnarray}\label{nlred0}
\vec w_{t}+V \vec w_{x_1}=0,\;\;\vec w= \left[
\begin{array}{c}
w\cr p\cr u\cr v
\end{array}
\right],\;\;\;
V=\left[
\begin{array}{cccc}
w&0&p&0\cr 
0&w&0&p\cr 
0&0&-w&-u\cr
0&0&-p&- v
\end{array}\right]
\end{eqnarray}
In general,
$4\times 4$ matrix of this system has $3$  different eigenvalues. Thus, this system is intermediate between
PDEs integrable by the generalized hodograph method \cite{ts1} (which requires four different eigenvalues) 
and method of characteristics for matrix equations \cite{SZ1} (which requires two different eigenvalues). 

We will show that solution space to the system (\ref{nl},\ref{nl_constr}) is implicitly described by the system of integral-algebraic equations which  mixtures   integral equation of 
the classical $\bar\partial$-problem  and algebraic equations typical for  the method of characteristics \cite{SZ1,ZS2}.
In particular cases, this integral-algebraic system becomes  system of 
algebraic equations, which is quite equivalent to the system  derived in \cite{Z4}, see also \cite{SZ1,ZS2}.
According to \cite{Z4}, this fact demonstrates that our nonlinear PDEs 
 possess solutions with wave profile breaking.

In the next section, Sec.\ref{Dressing}, we represent derivation of the systems (\ref{nl},\ref{nl_constr}) and its reduction (\ref{nlred}) by the dressing method. 
In Sec.\ref{Solutions} we describe  solution space to the system (\ref{nl},\ref{nl_constr}) and give some remarks on the construction of solutions to the eq.(\ref{nlred}).
Conclusions are given in Sec.\ref{Conclusions}.

\section{Dressing method: derivation  of nonlinear PDEs}
\label{Dressing}
 \subsection{Dressing and spectral functions}
 In this subsection we introduce basic functions and operators of the dressing algorithm.
 
 \paragraph{Homogeneous Fredholm equation and general form of the spectral system.}
We start with the following integral equation \cite{ZS2}:
\begin{eqnarray}\label{u1}
\int\limits_D \Psi(\lambda,\nu;x) U(\nu,\mu;x) d\nu\equiv\Psi(\lambda,\nu;x)*U(\nu,\mu;x) =0,
\end{eqnarray}
 where $\lambda$ and $\nu$ are complex (either scalar or vector) spectral parameters,  ''$*$'' means integration over some region $D$ of the  spectral parameter space, $x=(x_1,x_2,\dots,t_1,t_2\dots)$ is a set of all independent variables of nonlinear PDEs; $U$ is 
 $2Q\times Q$ matrix spectral function  depending on two spectral parameters; $\Psi$ is $Q\times 2Q$ dressing function and kernel of the integral operator.
Following the strategy of \cite{ZS2,Z4}, we assume that the general solution to  eq.(\ref{u1}) 
may be represented in the next form:
\begin{eqnarray}\label{U_sol}\label{U_sol1}
U(\lambda,\mu;x)=U^{(h)}(\lambda,\nu;x)* f(\nu,\mu;x),
\end{eqnarray}
where $f(\nu,\mu;x)$ is an arbitrary $Q\times Q$ matrix  function of arguments and $U^{(h)}$ is
a particular nontrivial solution to the homogeneous equation (\ref{u1}).
This assumption causes the unique linear  relation among any two independent solutions  $U^{(j)}$, $j=0,1,\dots$, of eq.(\ref{u1}):
\begin{eqnarray}\label{linU^i}
U^{(j)}(\lambda,\mu;x)=U^{(0)}(\lambda,\nu;x)*F^{(j)}(\nu,\mu;x),
\end{eqnarray}
where  $F^{(j)}$ are some $Q\times Q$ matrix  functions. 
As we shell see, all solutions $U^{(j)}$  are expressed in terms of the single solution  $U$  through some linear operators $L^{(j)}$, either differential or non-differential: 
$U^{(j)}(\lambda,\mu;x)=L^{(j)}(\lambda,\nu)*U(\nu,\mu;x)$.
 Thus, eqs.(\ref{linU^i}) represent the general form of the overdetermined compatible  system of  linear equations for the spectral function $U$ (general form of the spectral system). Besides, we will show in Sec.\ref{Dressing:PDE1} that  $F^{(j)}$ may be expressed in terms of $U$  using an external $Q\times 2Q$  dressing matrix function $G(\lambda,\mu;x)$, similar to \cite{ZS2,Z4}.
 \paragraph{$x$-dependence of the dressing function $\Psi$.}
We 
 introduce the  $Q\times Q$ matrix function  ${\cal{A}}(\lambda,\mu)$ and $2Q\times 2Q$ matrix function  $A(\lambda,\mu)$ (both functions are independent on $x$)
 by the following generalized commutation relation:
\begin{eqnarray}\label{Acom} {\cal{A}}(\lambda,\nu)*\Psi(\nu,\mu;x)=\Psi(\lambda,\nu;x)*A(\nu,\mu)
 \end{eqnarray}
and   define operators ${\cal{A}}^j$ and $A^j$ as follows: ${\cal{A}}^{j}=\underbrace{{\cal{A}}*\cdots*{\cal{A}}}_j$, $A^{j}=\underbrace{A*\cdots*A}_j$.
 Let $x$-dependence of $\Psi$ be given by the equation
 \begin{eqnarray}\label{x} \Psi_{t_m}(\lambda,\mu;x)+{\cal{A}}(\lambda,\nu)*\Psi_{x_m}(\nu,\mu;x) =0,
 \end{eqnarray}
which is compatible with eq.(\ref{Acom}). 

\paragraph{External dressing function $G$.}
We have to introduce an external dressing $Q\times 2Q$ matrix function $G(\lambda,\mu;x)$ which was mentioned above and  whose prescription will be explored in  Sec.\ref{Dressing:PDE1}.
Let $G$ be defined  
by the next compatible system of linear equations:
\begin{eqnarray}\label{G_com}
 &&
G(\lambda,\nu;x)*A(\nu,\mu) = \hat A(\lambda,\nu)*G(\nu,\mu;x)+ 
 \sum_{j=1}^2 H^{(j)}_1(\lambda;x) H^{(j)}_2(\mu;x)
 ,\\\label{G}\label{G_x}
&&
 G_{t_m}(\lambda,\mu;x) +G_{x_m}(\lambda,\nu;x)*A(\nu,\mu)=0,
 \end{eqnarray}
 where   $\hat A$ and $H^{(j)}_1$, $j=1,2$, are $Q\times Q$, 
 while $H^{(j)}_2$, $j=1,2$,  are $Q\times 2 Q$  matrix functions.
We refer to functions $H^{(j)}_i(\lambda;x)$,  $i,j=1,2$ as  external dressing functions as well.  
The  compatibility condition of eqs.(\ref{G_com}) and (\ref{G_x}) yields:
\begin{eqnarray}\label{comp00}\label{comp02}\label{G_com2}
&&
\sum_{j=1}^2 \left[ \Big(H^{(j)}_1(\lambda;x)\Big)_{t_m} H^{(j)}_2(\mu;x)+
 {H^{(j)}_1}_{x_m}(\lambda;x) H^{(j)}_2(\nu;x) *A(\nu,\mu)+\right.\\\nonumber
 &&\left.
H^{(j)}_1(\lambda;x)   \left({H^{(j)}_2}_{t_m}(\mu;x)+{H^{(j)}_2}_{x_m}(\nu;x) *A(\nu,\mu)\right)\right] =0,
\end{eqnarray}
which  admits the following solution:
\begin{eqnarray}\label{CH}\label{case01a}
&&
{H^{(j)}_1}_{t_m} ={H^{(j)}_1}_{x_j} =0,\;\;j=1,2,\\
\label{H22}\label{case01b}
&&
{H^{(j)}_2}_{t_m}+{H^{(j)}_2}_{x_m} *A  =0,\;\;j=1,2,
\end{eqnarray}
i.e. functions $H_1^{(j)}$ do not depend on $x$: $H_1^{(j)}(\lambda;x) \equiv H_1^{(j)}(\lambda)$, $j=1,2$. 
In order to derive PDEs different from the classical $S$-integrable systems we  require
\begin{eqnarray}\label{AAHH}
\hat A*H^{(1)}_1=0,\;\;\;\hat A*H^{(2)}_1\neq 0.
\end{eqnarray}

\subsection{Spectral system for $U(\lambda,\mu;x)$}
\label{Dressing:PDE1}
Now we are ready to derive the overdetermined linear system for the spectral function $U(\lambda,\mu;x)$, i.e. the spectral system.
Following the usual strategy of the dressing methods, we have to obtain set of 
 different solutions to the homogeneous  eq.(\ref{u1}) expressed in terms of functions $U$, $A$ and  $x$-derivatives of $U$. For this purpose we  apply 
${\cal{A}}^{m}*$ and 
$(\partial_{t_m} + {\cal{A}}*\partial_{x_m})$ to (\ref{u1}) and use eqs.(\ref{Acom},\ref{x}).
One gets 
\begin{eqnarray}\label{EE}
\Psi(\lambda,\nu;x)*E^{(j;m)}(\nu,\mu;x)=0,\;\;j=1,2,
\end{eqnarray}
where
\begin{eqnarray}\label{H2}
&&
E^{(1;m)}(\lambda,\mu;x)=A^{m}(\lambda,\nu)*
U(\nu,\mu;x),\\\nonumber
&&\label{H3}
E^{(2;m)}(\lambda,\mu;x)=U_{t_m}(\lambda,\mu;x) + A(\lambda,\nu) *U_{x_m}(\nu,\mu;x) .
\end{eqnarray}
Remember the eq.(\ref{linU^i}) relating any two different solutions of  the homogeneous equation (\ref{u1}). Let $U^{(0)}\equiv U$ in the eq.(\ref{linU^i})  and consider  $E^{(j;m)}$ as different solutions of the eq.(\ref{u1}).
Then eq.(\ref{linU^i}) yields:
\begin{eqnarray}\label{U_sp1}
&&
E^{(1;1)}(\lambda,\mu;x)=U(\lambda,\nu)*\tilde F(\nu,\mu;x),\;\; \Rightarrow 
\\\nonumber
&&
A(\lambda,\nu;x)*U(\nu,\mu;x) = U(\lambda,\nu;x)*\tilde F(\nu,\mu;x),
\\\label{U_sp2}
&&
E^{(2;m)}(\lambda,\mu;x)=U(\lambda,\nu)* F^{(m)}(\nu,\mu;x),\;\; \Rightarrow \\\nonumber
&&
U_{t_m}(\lambda,\mu;x) +  A(\lambda,\nu) *U_{x_m}(\nu,\mu;x) =
 U(\lambda,\nu;x)*F^{(m)}(\nu,\mu;x),\;\;m=1,2,\dots.
\end{eqnarray}
Eqs. (\ref{U_sp1},\ref{U_sp2}) represent a preliminary version of the
overdetermined linear system  for the spectral function $U(\lambda,\mu;x)$.

Recall that $U(\lambda,\mu;x)$ is not unique solution of the integral equation (\ref{u1}). To obtain uniqueness we have to introduce one more equation for the spectral function $U$. 
For instance, using the external dressing function $G$, we may write
 \begin{eqnarray}\label{condition}\label{sol_GU}
G(\lambda,\nu;x)*U(\nu,\mu;x)=I\delta(\lambda-\mu).
\end{eqnarray}
Now $U$ is {\it the unique} solution of the system (\ref{u1},\ref{condition}). In other words, the equation (\ref{condition}) fixes  function $f(\lambda,\mu;x)$ in the eq.(\ref{U_sol1}).
Applying  $G*$ to the eqs.(\ref{U_sp1},\ref{U_sp2}) and using eq.(\ref{condition}) one gets the expressions  for 
$\tilde F$ and $F^{(m)}$:
\begin{eqnarray}\label{F_def0}
\tilde F(\nu,\mu;x)&=&G(\lambda,\nu;x)*E^{(1;1)}(\nu,\mu;x)=\\\nonumber
&&\hat A(\lambda,\mu)+\sum_{j=1}^2
H^{(j)}_1(\lambda)  H^{(j)}_2(\nu;x)*U(\nu,\mu;x),\\\nonumber
F^{(m)}(\nu,\mu;x)&=&
G(\lambda,\nu;x)*E^{(2;m)}(\nu,\mu;x)= 
\\\nonumber
&&\sum_{j=1}^2
H^{(j)}_1(\lambda) \Big(H^{(j)}_2(\nu;x)*U( \nu,\mu;x)\Big)_{x_m}, \;\;m=1,2,\dots .
\end{eqnarray}
Although index $m$ may take any positive integer value (reflecting the existence of the hierarchy of commuting flows) we will take only two values $m=1,2$, which is enough to construct a complete system of nonlinear PDEs.
Unlikely the classical spectral systems, eqs.(\ref{U_sp1},\ref{U_sp2}) depend on two spectral parameters due to the spectral function $U(\lambda,\mu;x)$. 
However,  functions of single spectral parameter  appear in these equations naturally. These functions are following:
\begin{eqnarray}\label{WV}
&&
V^{(ji)}(\lambda;x)={U}(\lambda,\mu;x)*
\hat A^i(\mu,\nu)*H^{(j)}_1(\nu)
,\\\nonumber
&&
W^{(ji)}(\mu;x)= H^{(j)}_2(\lambda;x)*A^{i}(\lambda,\nu)*U(\nu,\mu;x),\;\;i,j=1,2.
\end{eqnarray}
All in all, substituting eqs.(\ref{F_def0}) and (\ref{WV}) into the eqs.(\ref{U_sp1},\ref{U_sp2}) one gets:
\begin{eqnarray}\label{U_sp30}\label{U_sp30a}
&&
\hspace{-1cm}
A(\lambda,\nu)*U(\nu,\mu;x) =U(\lambda,\nu;x)*\hat A(\nu,\mu) + \sum_{j=1}^2 V^{(j0)}(\lambda;x) W^{(j0)}(\mu;x),\\\label{U_sp30b}
&&\hspace{-1cm}
U_{t_m}(\lambda,\mu;x) + A(\lambda,\nu)*U_{x_m}(\nu,\mu;x) = 
  \sum_{j=1}^2 V^{(j0)}(\lambda;x) W^{(j0)}_{x_m}(\mu;x),\;\;m=1,2.
\end{eqnarray}
The non-classical type spectral system (\ref{U_sp30},\ref{U_sp30b}) depending on two spectral parameters $\lambda$ and $\mu$  
gives rise to the classical type spectral system for the spectral functions $V^{(j0)}(\lambda;x)$, $j=1,2$ with single spectral parameter. This system appears   
 after   applying  
$* H^{(k)}_1$, $k=1,2$ to the eqs.(\ref{U_sp30}) and using eqs.(\ref{AAHH}):
\begin{eqnarray}\label{U_sp21}
&&
A(\lambda,\nu;x)*V^{(10)}(\nu;x) =  
\sum_{j=1}^2 V^{(j0)}(\lambda;x)  w^{(j1;00)}(x),
\\\nonumber
&&
A(\lambda,\nu;x)* V^{(20)}(\nu;x) =   V^{(21)}(\lambda;x)+\sum_{j=1}^2
V^{(j0)}(\lambda;x)  w^{(j2;00)}(x),
\\\nonumber
&&
V^{(k0)}_{t_m}(\lambda;x) + A(\lambda,\nu) *V^{(k0)}_{x_m}(\nu;x) -
\sum_{j=1}^2 V^{(j0)}(\lambda;x)  w^{(jk;00)}_{x_m}(x) =0
,\;\;k,m=1,2,
\end{eqnarray}
where fields $w^{(ij;kn)}$ are defined as follows:
\begin{eqnarray}\label{ww}
&&
w^{(ij;kn)}=H^{(i)}_2*A^k*U*A^n*H^{(j)}_1.
\end{eqnarray}
The definition of fields (\ref{ww})  suggests us, for instance,  two following   reductions:
\begin{eqnarray}\label{red22}
1.&&\hat A^{n_0}* H^{(2)}_1=\sum_{i=1}^{n_0-1}\hat A^{i}*H^{(2)}_1r^{(i)} \;\;\Rightarrow
\;\;
w^{(j2;kn_0)}=\sum_{i=1}^{n_0-1}w^{(j2;ki)}r^{(i)},\;\;
\forall j,k.
\\
\label{red11}\label{redH2_1}
2. && H^{(2)}_2*A^{k_0}= \sum_{i=1}^{k_0-1} r^{(i)} H^{(2)}_2*A^{i} 
 \;\;\Rightarrow
\;\;\;
w^{(2j;k_0n)}= \sum_{i=1}^{k_0-1} r^{(i)} w^{(2j;in)}
,\;\forall j,n
\end{eqnarray}
where  $n_0$ and $k_0$ are any integer numbers and  $r^{(i)}$  are scalar parameters. 

\subsection{Nonlinear PDEs}
\label{Section:matr_g}\label{Char_Gen}
Applying $H^{(i)}_2*$, $i=1,2$  to the system (\ref{U_sp21}) we obtain the following  system of nonlinear PDEs:
\begin{eqnarray}\label{nl10}\label{nl10a}
&&
w^{(i1;10)} =  
\sum_{j=1}^2 w^{(ij;00)}  w^{(j1;00)},
\\\label{nl10b}
&&
 w^{(i2;10)} =   w^{(i2;01)}+\sum_{j=1}^2
w^{(ij;00)}  w^{(j2;00)},
\\\label{nl10c}
E^{(ik;00;m)}_1&:=&
 w^{(ik;00)}_{t_m} + w^{(ik;10)}_{x_m} -
\sum_{j=1}^2 w^{(ij;00)}  w^{(jk;00)}_{x_m} =0
,\;\;i,k,m=1,2.
\end{eqnarray}
Eliminating $w^{(i1;10)}$  from the eq.(\ref{nl10c}), $k=1$, using  eq.(\ref{nl10a}) one gets
\begin{eqnarray}\label{nl1}
&&
w^{(i1;00)}_{t_m} + 
\sum_{j=1}^2 w^{(ij;00)}_{x_m}  w^{(j1;00)} =0
,\;\;i,m=1,2.
\end{eqnarray}
Eq.(\ref{nl10c}), $k=2$ in view of eq.(\ref{nl10b}) may be given another form,
\begin{eqnarray}\label{nl10d}
E^{(i2;00;m)}_2:=w^{(i2;00)}_{t_m}+w^{(i2;01)}_{x_m} +
\sum_{j=1}^2 w^{(ij;00)}_{x_m}  w^{(j2;00)} =0
,\;\;i,m=1,2,
\end{eqnarray}
which is convenient for imposing the reduction (\ref{red22}).
Now the complete system of nonlinear PDEs is represented by the eqs.(\ref{nl1})
and the following combination of eqs.(\ref{nl10c}), $k=2$:   $({E^{(i2;00;1)}_1})_{x_2}-({E^{(i2;00;2)}_1})_{x_1}$. 
Introducing   new dependent and independent variables
\begin{eqnarray}
w=w^{(11;00)},\;\;\;p=w^{(21;00)},\;\;\;u=w^{(12;00)},\;\;\;v=w^{(22;00)},\;\;\;t=t_1,\;\;\;x_3=t_2,
\end{eqnarray}
we  end up with the system (\ref{nl},\ref{nl_constr}).

\paragraph{Reductions.} Reduction (\ref{red22}) yelds quasilinear  matrix first order PDEs integrable by the method of characteristics \cite{SZ1}. Similar PDEs have been considered in \cite{Z4} as  lower 
dimensional reductions of appropriate Self-dual type $S$-integrable PDEs. It was shown  that such lower dimensional PDEs  generate solutions with wave profile breaking.
A new type of reductions is represented by the eq.(\ref{red11}). 
In the simplest example    $k_0=1$, the eqs.(\ref{nl10c})  and  (\ref{nl1}) with $m=1$ yield the system (\ref{nlred}).

\section{Implicit description of solutions  to nonlinear PDEs }
\label{Solutions}
\label{sol_PDE1}
We 
introduce the next block-matrix representation of the functions, $\forall \; i,j$:
\begin{eqnarray}\label{vec}
\Psi(\lambda,\mu;x)=[
\psi_{0}(\lambda,\mu;x) \;\;
\psi_{1}(\lambda,\mu;x)],&&
U(\lambda,\mu;x)=\left[\begin{array}{c}
u_{0}(\lambda,\mu;x) \cr
u_{1}(\lambda,\mu;x)\end{array}\right],\\\nonumber 
V^{(ij)}(\lambda;x)=\left[\begin{array}{c}
v^{(ij)}_{0}(\lambda;x)\cr 
v^{(ij)}_{1}(\lambda;x)\end{array}\right],&&
W^{(ij)}(\mu;x)=
w^{(ij)}_0(\mu;x),
\\\nonumber
G(\lambda,\mu;x)=[
g_{0}(\lambda,\mu;x)\;\; g_{1}(\lambda,\mu;x)],&&
H^{(i)}_2(\lambda;x)=[
h^{(i)}_{20}(\lambda;x) \;\;
h^{(i)}_{21}(\lambda;x)],
\\\nonumber
H^{(i)}_1(\lambda;x)= h^{(i)}_1(\lambda;x).&&
\end{eqnarray}
Any function in the RHS of the formulae (\ref{vec}) is  $Q\times Q$ matrix function, so that $\Psi$, $G$ and $H^{(i)}_2$ are $Q\times 2 Q$, $U$ and $V^{(ij)}$ are $2 Q\times Q$,
$W^{(ij)}$ and $H^{(i)}_1$ are $Q\times Q$ matrix functions.
We also have to fix the  functions  ${\cal{A}}(\lambda,\mu)$, $A(\lambda,\mu)$ and $\hat A(\lambda,\mu)$:
\begin{eqnarray}\label{AAA0}
{\cal{A}}(\lambda,\mu) =\hat A(\lambda,\mu)= \lambda
\delta(\lambda-\mu)I
,\;\;\;
A(\lambda,\mu)=
\lambda \delta(\lambda-\mu)I_2, 
\end{eqnarray}
where $I$ and $I_2$  are  $Q\times Q$ and $2Q\times 2Q$ identity matrices respectively.
 Eq.(\ref{Acom}) suggests us the following structure  of   $\Psi$:
\begin{eqnarray}\label{hatPsi}
 \Psi(\lambda,\mu;x) =\hat \Psi(\lambda;x) \delta(\lambda-\mu),\;\;
\hat\Psi(\lambda;x)=[\hat \psi_{0}(\lambda;x)\;\; \hat \psi_{1}(\lambda;x)],
\end{eqnarray}
where $\hat \psi_{i}$, $i=0,1$, are $Q\times Q$ matrix functions.

 The main feature of the spectral system (\ref{U_sp30},\ref{U_sp30b})  is the presence of the spectral equation which has no derivatives with respect to $x$, see eq.(\ref{U_sp30a}),
which suggests us the next representation  for $U$:
\begin{eqnarray}\label{sol_U}
&&
U(\lambda,\mu;x)=\frac{\sum_{j=1}^2 V^{(j0)}(\lambda;x) W^{(j0)}(\mu;x)
}{\lambda-\mu} + 
 U_0(\lambda;x)\delta(\lambda-\mu),
\\\nonumber
&&
U_0(\lambda;x)=\left[\begin{array}{c}u_{00}(\lambda;x)\cr
 u_{01}(\lambda;x)\end{array}\right],
\end{eqnarray}
where $u_{0i}$, $i=0,1$, are $Q\times Q$  matrix functions.
 Remark, that eq.(\ref{U_sp30a}) after applying $* H^{(1)}_1$ yields:
\begin{eqnarray}\label{tVV}
 V^{(10)}(\lambda;x) (\lambda I  - w^{(11;00)}(x))=V^{(20)}(\lambda;x)w^{(21;00)}(x) 
,\;\;
\end{eqnarray}
which relates spectral functions $V^{(10)}$ and $V^{(20)}$. Let us write this relation for the case of diagonalizable $w^{(11;00)}$, i.e.
\begin{eqnarray}\label{PEP}\label{PEPa}
&&
w^{(11;00)}(x)=P(x) E(x) P^{-1}(x),\\\label{PEPb}
&&
P_{\alpha\alpha}=1,
\end{eqnarray}
where $E$ is the diagonal matrix of eigenvalues, $P$ is the matrix of eigenvectors with normalization (\ref{PEPb}).
Then, multiplying eq.(\ref{tVV}) by $P(\lambda I  -E(x))^{-1}$ from the right one gets
\begin{eqnarray}\label{hatV}
&&
V^{(10)}(\lambda;x) P =  V^{(20)}(\lambda;x)w^{(21)}(x)(\lambda I  -E(x))^{-1} + \hat V(\lambda;x) \delta(\lambda I - E(x)), \\\nonumber
&&
\hat V =\left[
\begin{array}{c}
\hat v_0\cr\hat v_1
\end{array}
\right],\;\;w^{(21)}=w^{(21;00)} P.
\end{eqnarray}

Similarly, eq.(\ref{G_com}) yelds
\begin{eqnarray}\label{sol_G}
&&
G(\lambda,\mu;x)=-\sum_{j=1}^2 \frac{H^{(j)}_1(\lambda;x) H^{(j)}_2(\mu;x)
}{\lambda-\mu}+  G_0(\lambda;x)\delta(\lambda-\mu),\\\nonumber
&&
G_0(\lambda;x)=[g_{00}(\lambda;x)\;\; g_{01}(\lambda;x)],
\end{eqnarray}
where $g_{0i}$, $i=0,1$, are $Q\times Q$  matrix functions and $G_0(\lambda;x)\delta(\lambda-\mu) $ is a solution of the eq.(\ref{G_x}).

Next, after substitution  the eqs.(\ref{sol_U}) and (\ref{sol_G}) into the eq.(\ref{sol_GU}), one gets:
\begin{eqnarray}\label{EEE}
&&
\frac{1}{\lambda-\mu}\sum_{j=1}^2
\left[ E_{1j}(\lambda;x) W^{(j0)}(\mu;x)  +  H^{(j)}_1(\lambda;x) E_{2j}(\mu;x)
\right]  +
\\\nonumber
&&
\hspace{2cm} G_0(\lambda;x) U_0(\lambda;x)\delta(\lambda-\mu) =
 I\delta(\lambda-\mu),
 \\\nonumber
&&
E_{1j}(\lambda;x)=G_0(\lambda;x) V^{(j0)}(\lambda;x)-
\\\nonumber
&&
\hspace{2cm}\int\limits_D d\nu \frac{ \sum_{i=1}^2
H^{(i)}_1(\lambda;x) H^{(i)}_2(\nu;x)  V^{(j0)}(\nu;x)}{\lambda-\nu}  - H^{(j)}_1(\lambda;x),
\\\nonumber
&&
E_{2j}(\mu;x)=-\int\limits_D d\nu\frac{
  H^{(j)}_2(\nu;x)  \sum_{i=1}^2V^{(i0)}(\nu;x)   W^{(i0)}(\mu;x)
 }{\nu-\mu} -  
 \\\nonumber
 &&
\hspace{2cm} H^{(j)}_2(\mu;x)   U_0(\mu;x) +  W^{(j0)}(\mu;x).
\end{eqnarray}
The eq.(\ref{EEE}) must be identity for any $\lambda$ and $\mu$. Thus, it must be splited into the following set of equations:
\begin{eqnarray}\label{sol_G_0}
&&
 G_0(\lambda;x)  U_0(\lambda;x) = I,
\\\label{E10}
&&
E_{1j}(\lambda;x)=0,\\\label{E20}
&&
E_{2j}(\mu;x)=0.
\end{eqnarray} 
The last terms in the expressions  $E_{2j}$ have been introduced in order to 
eqs.(\ref{E20}) coincide with eq.(\ref{sol_U}) after applying $H^{(j)}_2*$  to eq.(\ref{sol_U}), which is necessary condition. The last terms in the expressions for $E_{1j}$ are needed to compensate the last terms of $E_{2j}$  in the eq.(\ref{EEE}).

The system (\ref{E20}) may be viewed as  $2Q^2$ scalar equations for $ 2Q^2$ elements of the matrix functions $ W^{(j0)}$, $j=1,2$, i.e. $W^{(j0)}$ are completely defined. 
However, eqs.(\ref{sol_G_0},\ref{E10}) are not a complete system for $U_0$ and $V^{(j0)}$, $j=1,2$. In fact,  eq.(\ref{sol_G_0}) represents $Q^2$ scalar equations for $2 Q^2$ elements of the matrix function $ U_0$.
Similarly, eq.(\ref{E10})  represents $2Q^2$ scalar equations for $4 Q^2$ elements of the matrix functions $  V^{(j0)}$, $j=1,2$. Thus, both eq.(\ref{sol_G_0})  and  eq.(\ref{E10}) are underdetermined systems. 

The rest of equations for the elements of $  V^{(j0)}$ and $ U_0$ follows from  the eq.(\ref{u1})  after substitution  the eq.(\ref{hatPsi}) for $ \Psi$ and  the eq.(\ref{sol_U}) for $ U$:
\begin{eqnarray}\label{PsiVW}
 \hat \Psi(\lambda;x) \frac{ \sum_{j=1}^2V^{(j0)}(\lambda;x) W^{(j0)}(\mu;x)
 }{\lambda-\mu}+ \hat \Psi(\lambda;x)  U_0(\lambda;x) \delta(\lambda-\mu) =0. 
\end{eqnarray}
Since eq.(\ref{PsiVW}) must be identity for any $\lambda$ and $\mu$, it 
is equivalent to the following  equations for $ V^{(j0)}$ and $ U_0$:
\begin{eqnarray}\label{PsiV}
&&
\hat\Psi(\lambda;x) V^{(j0)}(\lambda;x) =0,\;\;j=1,2 \;\;\overset{{(\ref{hatV})}}{\Rightarrow} \;\; \hat\Psi(\lambda;x) \hat V(\lambda;x) = 0,\\\label{PsiU0}
&&
\hat\Psi(\lambda;x) U_0(\lambda;x)=0
\end{eqnarray}
Each of these matrix equations represents $Q^2$ scalar equations 
for $2Q^2$ elements of one of the matrix functions  $ V^{(j0)}$, $j=1,2$ and 
$U_0$. Thus, the system 
(\ref{sol_G_0}-\ref{E20},\ref{PsiV},\ref{PsiU0}) is 
the complete system for elements of $  V^{(j0)}(\lambda;x)$,  $
W^{(j0)}(\mu;x)$, $j=1,2$ and $  U_0(\lambda;x)$. 
Having these functions,  the spectral function $ U(\lambda,\mu;x)$ may be constructed using the formula (\ref{sol_U}).

It is simple to observe, that essentially important for construction of $w^{(ij;00)}=H^{(i)}_2*V^{(j0)}$ are eqs.(\ref{E10}) and (\ref{PsiV}) defining $V^{(j0)}$.
The following Proposition is valid.

\begin{theorem}
 If reductions (\ref{red22}) and (\ref{red11})  have not been involved into consideration, then
 \begin{enumerate}
 \item
Eqs.(\ref{E10}) and (\ref{PsiV}) are equivalent to the next system of 
integral-algebraic equations:
\begin{eqnarray}\label{inter_2}\label{int_dbar}\label{inter_4}
 &&\hspace{-1cm} v^{(20)}_0(\lambda;x)-
\int\limits_{D} d\nu \frac{ 
 \chi^{(2)}(\nu;x)  v^{(20)}_0(\nu;x)}{\lambda-\nu} 
 +
 \delta(\lambda)w_0(x)  =
 I,
\\\nonumber
&&
\hspace{2cm} w_0(x) =\int\limits_{D} d\nu\;\frac{ \chi^{(1)}(\nu;x) v^{(20)}_0(\nu;x)}{\nu}
,\\\label{inter_3}
&&\hspace{-1cm}
  \int\limits_{D} d\nu\;
  \chi^{(1)}(\nu;x)  v^{(20)}_0(\nu;x) w^{(21;00)}(x) (\nu I-w^{(11;00)}(x))^{-1}
 = w^{(11;00)}(x),
\end{eqnarray}
where
\begin{eqnarray}
 \label{dbar_sol_Psi}
\chi^{(j)}(\lambda;x)=
\int_{\RR^{N}}\limits dq\;
 e^{I\sum\limits_{m=1}^2 q_m (x_m - 
\lambda  t_m)
}\chi^{(j)}_{0}(\lambda,q),\;\;j=1,2.
\end{eqnarray}
Here $\chi^{(j)}_0$, $j=1,2$  are  arbitrary $Q\times Q$ matrix functions, $q=(q_1,\dots,q_N)$, parameters $q_i$ are complex in general.
\item
Expressions for   fields $w^{(j2;00)}$, $j=1,2$,  and  $w^{(21;00)}$ follow from the definition (\ref{ww}):

\begin{subequations} \label{w^21}
\begin{gather}
\label{w^21b}
w^{(21;00)}(x)=\chi^{(2)}(\lambda;x)*v^{(10)}_0(\lambda;x)
,\\
\label{w^21a}
w^{(j2;00)}(x)=\chi^{(j)}(\lambda;x)*v^{(20)}_0(\lambda;x),\;\;j=1,2,
\end{gather}
\end{subequations}
where
\begin{eqnarray}
\label{w^21c}
\hspace{1cm}
v^{(10)}_0(\lambda;x)  =  v^{(20)}_0(\lambda;x)w^{(21)}(x)(\lambda I  -w^{(11;00)}(x))^{-1} \end{eqnarray}
\item
Eqs.(\ref{inter_2},\ref{inter_3},\ref{w^21}) represent the complete system of integral-algebraic equations which defines fields $w^{(ij;00)}$, $i,j=1,2$.
\end{enumerate}
\end{theorem}
\begin{proof}
 To satisfy the eqs.(\ref{case01a}) and (\ref{AAHH}) with $\hat A$ given by the first of eqs.(\ref{AAA0}) we take 
\begin{eqnarray}\label{class_H1}
H^{(1)}_1(\lambda;x)=\delta(\lambda)I,  \;\;\;H_1^{(2)}(\lambda)=I.
\end{eqnarray}
 Note that this form of $H_1^{(2)}$ may be used 
unless reduction (\ref{red22}) is  involved. 
Then eqs.(\ref{PsiV}) and (\ref{PsiU0}) yield respectively
\begin{eqnarray}\label{vu} \label{vua} 
&&
v^{(j0)}_1(\lambda;x)=-\hat\psi^{-1}_{1}(\lambda;x)
\hat\psi_0(\lambda;x)  v^{(j0)}_0(\lambda;x),\;\;j=1,2,\\\label{vub}
&&
\hat v_1(\lambda;x)=-\hat\psi^{-1}_{1}(\lambda;x)
\hat\psi_0(\lambda;x)  \hat v_0(\lambda;x),\\\label{vuc}
&&
u_{01}(\lambda;x)=-\hat\psi^{-1}_{1}(\lambda;x)\hat\psi_0(\lambda;x)  u_{00}(\lambda;x) .
\end{eqnarray}
Thus, eq.(\ref{E10}), $j=2$, gets the next form:
\begin{eqnarray}\label{inter_1}
 &&
 \phi(\lambda;x) v^{(20)}_0(\lambda;x)-
\int\limits_D d\nu \frac{ \Big(
 \chi^{(2)}(\nu;x)+\delta(\lambda) \chi^{(1)}(\nu;x)  \Big)v^{(20)}_0(\nu;x)}{\lambda-\nu} 
  = 
 I,
\end{eqnarray}
where
\begin{eqnarray}
&&
\phi(\lambda;x)= g_{00}(\lambda;x)- g_{01}(\lambda;x)
\hat\psi^{-1}_{1}(\lambda;x)\hat\psi_0(\lambda;x),
\\\nonumber
&&
\chi^{(j)}(\lambda;x)= h^{(j)}_{20}(\lambda;x)- h^{(j)}_{21}(\lambda;x)\hat\psi^{-1}_{1}(\lambda;x)\hat\psi_0(\lambda;x),\;\;j=1,2.
\end{eqnarray}
 Requiring  $\phi(\lambda;x) =I$ we get eq.(\ref{int_dbar}).
 Function $\chi^{(2)}$ must provide uniqueness of $v^{(20)}_0$ as a solution of eq.(\ref{inter_2}). 
 Classical $\bar\partial$-problem for Pohlmeyer equation corresponds to $\chi^{(1)}=0$ (and, as a consequence, $w_0=0$) \cite{ZM0,Z0}.
 
Multiplying eq.(\ref{E10}), $j=1$,  by $P$ from the right,  substituting eq.(\ref{hatV}) for $V^{(10)} P$, using eqs.(\ref{vua}) with $j=1$ and  eq.(\ref{int_dbar}) for 
$v^{(20)}_0(\lambda;x)$ we obtain:
  \begin{eqnarray}
&&\hspace{-1cm}
   \hat v_0(\lambda;x)\delta(\lambda I-E) +
  \delta(\lambda)\left[ 
  \int\limits_{D} d\nu\;
 \chi^{(1)}(\nu;x)  v^{(20)}_0(\nu;x) w^{(21)}(x) (\nu I-E)^{-1}E^{-1}+\right.
  \\\nonumber
  &&\left.
  \int\limits_{D} d\nu\;
  \frac{\chi^{(1)}(\nu;x) \hat v_0(\nu;x)}{\nu} \delta(\nu I -E) 
 - P(x)
   \right]=0,
  \end{eqnarray}
  which is equivalent to the next pair of equations:
  \begin{eqnarray}\label{w^11_a}
  &&
   (\hat v_0(E_\beta;x))_{\alpha\beta} =0,\;\;\alpha,\beta=1,\dots,Q,\\\label{w^11_b}
  \label{w^11_a3}
 && \int\limits_{D} d\nu\;
  \chi^{(1)}(\nu;x) v^{(20)}_0(\nu;x) w^{(21)}(x) (\nu I-E(x))^{-1}E^{-1}(x)
 = P(x) 
  .
  \end{eqnarray} 
The matrix  equation (\ref{w^11_b}) is a system of scalar equations
for the elements of  $E$ and $P$.  We may replace the matrices $E$ and $P$ by $w^{(11;00)}$
in the eq.(\ref{w^11_a3}). For this purpose we  multiply eq.(\ref{w^11_a3}) by
$E P$ from the right resulting in the eq.(\ref{inter_3}).

Thus, we have derived  an integral equation (\ref{inter_2}) for $v^{(20)}_0$ and algebraic eq.(\ref{inter_3}) as an equation  relating fields $w^{(i1;00)}$, $i=1,2$. 
One more equation relating  these two fields  follows from the eqs.(\ref{ww}) (with $i=2$, $j=1$, $k=n=0$) and may be written as eq.(\ref{w^21b}).  Eq.(\ref{w^21c}) follows from the eq.(\ref{hatV}) after applying $P^{-1}$ from the right and using eqs.(\ref{vua}) and (\ref{w^11_a}).
Two more fields $w^{(i2;00)}$, $i=1,2$, 
may be calculated using  eqs.(\ref{ww}) with $i=1,2$, $j=2$, $k=n=0$, see eqs.(\ref{w^21a}).

Functions $\chi^{(j)}$, $j=1,2$,  satisfy the linear PDEs which follow from the linear PDEs for  the functions  $\hat \psi_i$
and $h^{(j)}_{2i}$, $i=0,1$. These PDEs are 
eqs.(\ref{x}) and (\ref{case01b}), which may be written as a single linear PDE
\begin{eqnarray}\label{psi_diff0}
&&
\varphi_{t_m}(\lambda;x) + \lambda
  \varphi_{x_m} (\lambda;x) 
 =0,\;\;m=1,2,
\end{eqnarray}
where $\varphi$ is one of the functions  $\hat \psi_i$ or $h^{(j)}_{2i}$, $i=0,1$ $j=1,2$.
 This means that $\chi^{(j)}$  are solutions of the same PDE as well, i.e. $\chi^{(j)}$
  may be written in the form (\ref{dbar_sol_Psi}).
\end{proof}

{\paragraph{Reductions.}
There is a remarkable sub-manifold of particular solutions corresponding to the reduction (\ref{red22}). System of integral-algebraic equations (\ref{inter_4},\ref{inter_3},\ref{w^21}) will be replaced by the system of algebraic equations. In the simplest case $n_0=1$, $A(\lambda,\mu)=\lambda\delta(\lambda-\mu) $,
$\hat A(\lambda,\mu)=\lambda (\lambda-a)\delta(\lambda-\mu)$,
  $H^{(1)}_1(\lambda)=\delta(\lambda) I$, 
  $H^{(2)}_1(\lambda)=\delta(\lambda-a) I$, $a=const$. 
  Then eq.(\ref{U_sp30a}) after applying $*H_1^{(1)}$ and $*H_1^{(2)}$ yields:
\begin{eqnarray}
\lambda{\bf V}(\lambda;x) = {\bf V}(\lambda;x) w(x),\;\;\;
{\bf V}=[V^{(1)} \;\;\;V^{(2)}],\;\; w=\left[
\begin{array}{cc}
w^{(11;00)} & w^{(12;00)}\cr
w^{(21;00)} & w^{(22;00)}
\end{array}
\right].
\end{eqnarray}
In result we will get the next algebraic system implicitly describing some family of solutions:
\begin{eqnarray}
w_{\alpha\beta}=\sum_{\gamma=1}^Q\left[ F_{\alpha\gamma}(x_1-w t_1,x_2-wt_2)\right]_{\gamma\beta},\;\;\alpha,\beta=1,\dots,Q,
\end{eqnarray}
which may be derived  by the algebraic method \cite{SZ1}.
Here $F(z_1,z_2)$ is arbitrary $Q\times Q$ matrix function.

The second reduction, eq.(\ref{red11}), is assotiated with more complicated form of functions $A(\lambda,\mu)$, $\hat A(\lambda,\mu)$, $H_1^{(1)}(\lambda)$ and $H_2^{(2)}(\lambda)$. The possible choice might be two-component
spectral parameter $\lambda=(\lambda_1,\lambda_2)$, $A(\lambda,\mu)=\lambda_1 \delta(\lambda_1-\mu_1)\delta(\lambda_2-\mu_2)$, $\hat A(\lambda,\mu)=\lambda_2 \delta(\lambda_1-\mu_1)\delta(\lambda_2-\mu_2)$, $H_1^{(1)}(\lambda)=\delta(\lambda_2)I$, 
$H_2^{(2)}(\lambda;x)=\delta(\lambda_1)I$. However, we postpone the detailed study of this reduction.

\section{Conclusions}
\label{Conclusions}

We represent a simplest example of nonlinear PDEs which may be treated by a version of the dressing method and admits reductions to Self-dual type $S$-integrable PDEs as well as to PDEs integrable by the method of characteristics. 
Remember, that similar  joining of $S$- and $C$-integrable models (see \cite{Calogero} for definition of $C$-integrability) has been represented in  \cite{Z}. In \cite{ZS2} we have considered a version of the dressing method joining  $C$-integrability and integrability by the method of characteristics. It is remarkable, that system (\ref{nl},\ref{nl_constr}) admits reduction to the system (\ref{nlred}), which may not be referred to neither PDEs integrable by the generalized hodograph method nor to PDEs integrable by the  method of characteristics.
All these examples demonstrate flexibility of the dressing method.

\begin{acknowledgments}
The work was supported by  RFBR grants 04-01-00508, 06-01-90840, 06-01-92053 
and grant NS 7550.2006.2.

\end{acknowledgments}

\end{document}